\documentclass[]{spie}  

\usepackage{wrapfig}
\usepackage{amsmath,amsfonts,amssymb}
\usepackage{graphicx}
\usepackage[colorlinks=true, allcolors=blue]{hyperref}
\usepackage{subfig}
\usepackage{float}
\usepackage{siunitx}
\usepackage{mhchem}
\newcommand{\xsig}{$X^{1}\Sigma^{+}_{g}$}
\newcommand{\bsig}{$B\,^{1}\Sigma^{+}_{u}$}
\newcommand{\cpi}{{$C\,^{1}\Pi_{u}$}}
\newcommand{\lam}{$\lambda$}

\title{Lyman-alpha Filter Prototype to Enable Astronomical Photometry in the Lyman Ultraviolet}

\author[1]{Isu Ravi}
\author[1]{Stephan R. McCandliss}
\author{Russell Pelton}
\affil{Center for Astrophysical Sciences\\ Department of Physics and Astronomy, Johns Hopkins University, 3400 North Charles Street, Baltimore, MD 21218, USA}

\begin{document} 
\maketitle

\begin{abstract}
Observations of astronomical objects in the far ultraviolet (FUV wavelengths span 900-\SI{1800}{\angstrom}) from earth's orbit has been impeded due to bright Lyman-$\alpha$ geocoronal emission.  The Johns Hopkins Rocket Group is developing a hydrogen absorption cell that would act as a narrow band Lyman-$\alpha$ rejection filter to enable space-based photometric observation in bandpasses that span over the  Lyman ultraviolet region shortward of the geocoronal line.  While this technology has been applied to various planetary missions with single element photomultiplier detectors it has yet to be used on near earth orbiting satellites with a multi-element detector.  We are working to develop a cell that could be easily incorporated into future Lyman ultraviolet missions.  The prototype cell is a low-pressure ($\sim$ few torr) chamber sealed between a pair of MgF$_{2}$ windows allowing transmission down to 1150 Å.  It is filled with molecular hydrogen that is converted to its neutral atomic form in the presence of a hot tungsten filament, which allows for the absorption of the Lyman-$\alpha$ photons.  Molecular hydrogen is stored in a fully saturated non-evaporable getter module (St707TM), which allows the cell pressure to be increased under a modest application of heat (a 20 degree rise from room temperature has produced a rise in pressure from 0.6 to 10 torr).  Testing is now underway using a vacuum ultraviolet monochromator to characterize the cell optical depth to Lyman-$\alpha$ photons as functions of pressure and tungsten filament current.  We will present these results, along with a discussion of enabled science in broadband photometric applications.

\end{abstract}


\section{INTRODUCTION}

Atomic hydrogen (\ion{H}{1}) is the most ubiquitous element in the universe, it exists in stars, galaxies, dust clouds, and even in “empty” space.  On Earth it is rarely found in its atomic state, but because of its simplistic makeup of a single proton and electron, it was the first atom modeled and understood.  In the realm of ultraviolet (UV) astronomy, hydrogen is observable through its Lyman series transitions, i.e., transitions from electronic states of n$\geq$2 down to the ground state, which spans a bandwidth of 912-\SI{1216}{\angstrom} often called the Lyman UV (LUV).  The strongest of these is the \ion{H}{1} ${\lambda}$1216 Lyman-$\alpha$ line  (Ly$\alpha$), which is the transition from the first excited state to the ground state.

\begin{figure} [t]
    \centering
    \subfloat[\centering]{{\includegraphics[height=7cm, width=8cm]{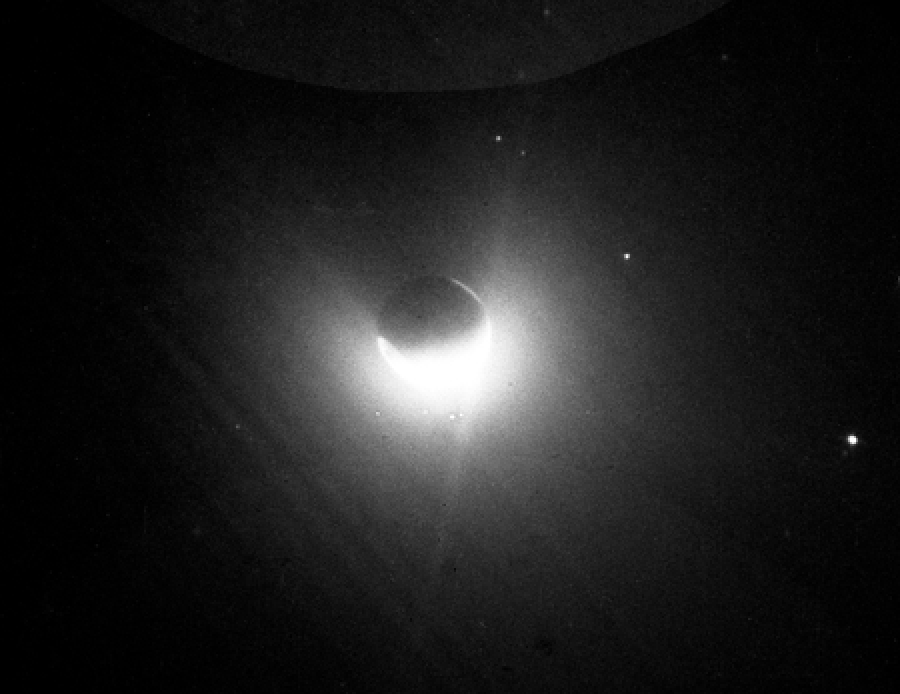} }}%
    \qquad
    \subfloat[\centering]{{\includegraphics[height=7cm, width=8cm]{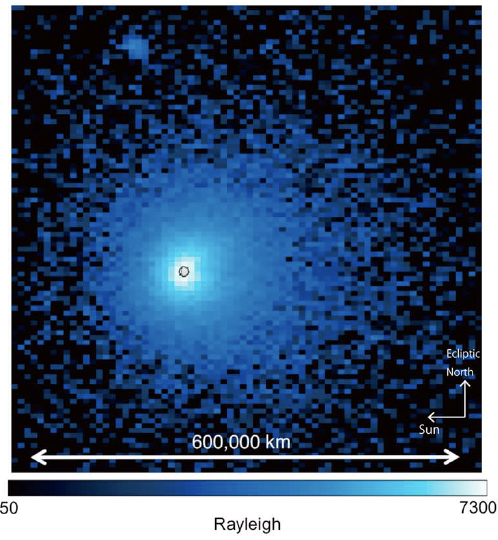} }}%
    \caption{\textbf{(a)} First image taken of geocorona from the moon by the Apollo 16 mission astronauts using imager built by George Carruthers. \cite{Carruthers76}  \textbf{(b)} Photometric image of the Earth's geocorona taken from the PROCYON spacecraft by Kameda et al. (2017). \cite{Kameda17}}%
    \label{f1}%
\end{figure}

Earth’s exosphere, the final collisionless boundary between our atmosphere and space, consists mainly of \ion{H}{1}.  As water vapor rises up through the atmosphere it is photodissociated into \ce{OH} and \ce{H}  by UV radiation from the sun (2390-\SI{1650}{\angstrom}) forming a hydrogen rich layer that resonantly scatters solar Ly$\alpha$ radiation\cite{Brinkmann69}.  The existence of a bright hydrogen emitting layer, called the geocorona, has been known since the early days of UV astronomy and was first imaged by the Apollo 16 mission seen in Figure~\ref{f1}a\cite{Carruthers76}.  It has been recently observed to exist well beyond the orbit of the moon, shown in Figure~\ref{f1}b.\cite{Baliukin18}.  Given the intensity of the geocorona, the existence of this extremely strong emission has completely stymied the development of photometric observations using instruments whose bandpass is sensitive to Ly$\alpha$.  For the field of UV astronomy, the geocorona is a massive hurdle that limits the commensurate depth of wide-field photometric surveys in comparison to other bandpasses.

Here we outline the science utilizing an edge filter photometric system that would be enabled if the geocorona no longer posed a problem.  We then describe the hydrogen absorption cell that will overcome this issue by filtering out geocoronal Ly$\alpha$.  We also discuss steps toward perfecting and testing this instrument's incorporation into future space based missions.

\section{LUV Observations of the Interstellar Radiation Field}

\begin{figure}[t] 
    \centering
    \subfloat[\centering]{{\includegraphics[width=.49\textwidth, viewport= 0in 0in 6.25in 5in, height=6cm]{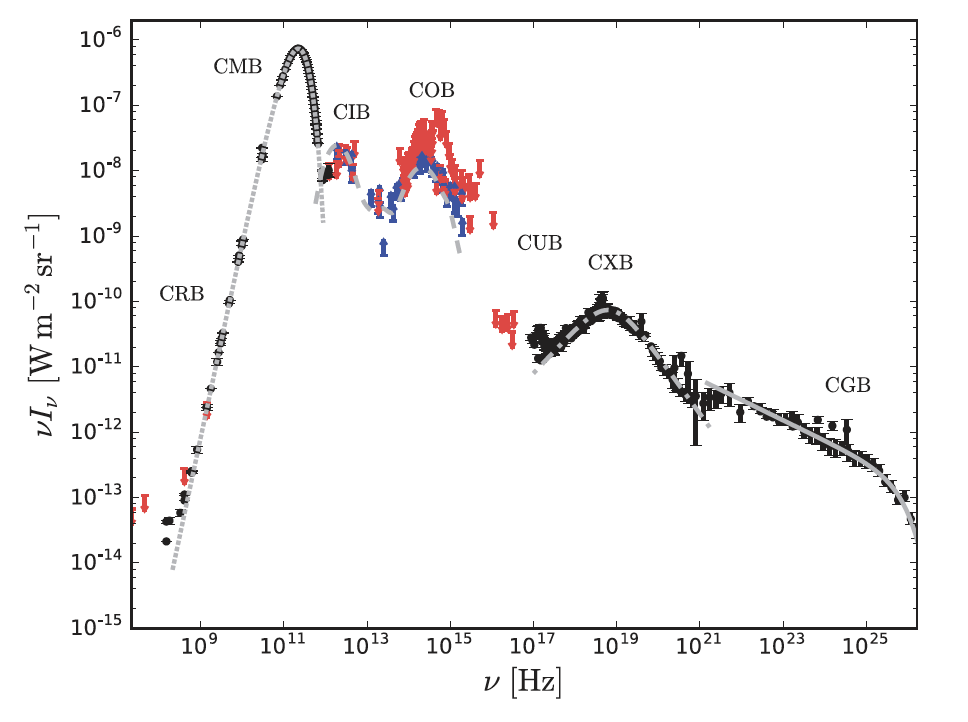} }}%
    \qquad
    \subfloat[\centering]{{\includegraphics[width=.4\textwidth, viewport= .2in 0in 6.8in 5in, height=6.5cm]{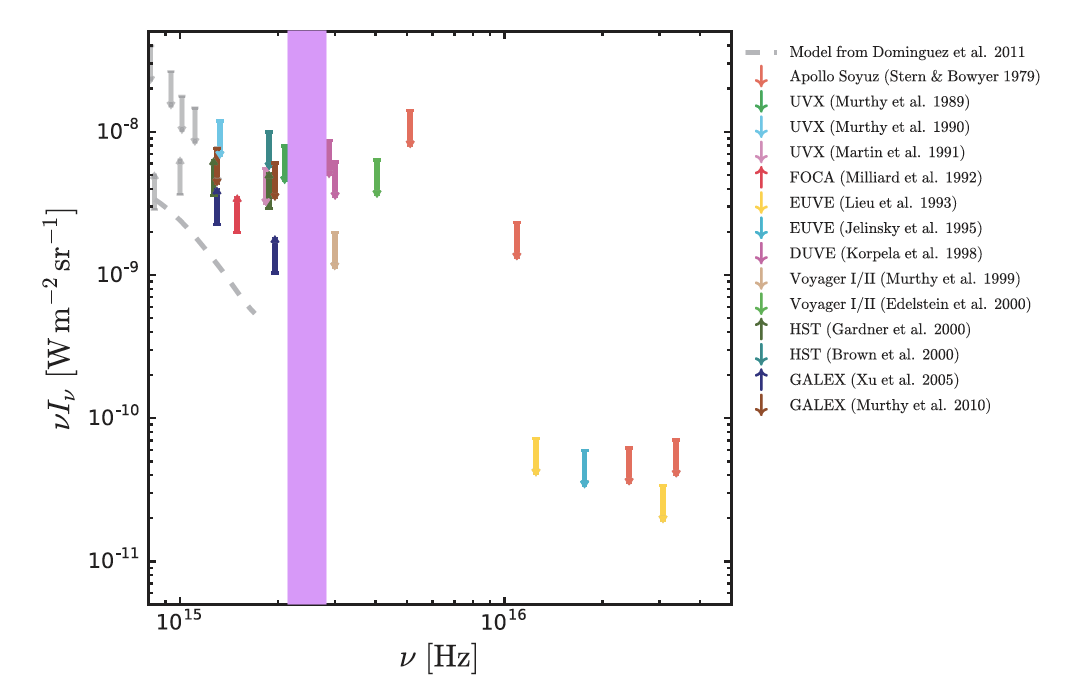} }}%
    \caption{\textbf{(a)} Measurements of the cosmic background in all frequency ranges.  \textbf{(b)} Compilation of the cosmic UV background measurements taken from various missions.  The LUV spectrum is highlighted in purple where no measurement exists.  Both figures were compiled by Hill et al. (2018)\cite{Hill18}}%
    \label{f2}%
\end{figure}

LUV photons play a major role in defining the chemistry of the Galaxy through molecular photodissociation processes such as the excitation and destruction of \ce{H2} in the interstellar medium\cite{Heays17}.  The interstellar radiation field (ISRF)  in the LUV has three components, dust scattered light from O and B stars, molecular hydrogen fluorescence, and the extragalactic background radiation\cite{Bowyer91, Henry91, Murthy10}.
 Although theoretical models exist for the ISRF, at present there have been no successful wide-field measurements of its intensity in the LUV bandpass, as shown in Figure~\ref{f2}.

\paragraph{Dust Scattered Starlight}
Dust scattered starlight is the main contributor to the ISRF.  Due to the brightness of the geocoronal Ly$\alpha$, at present, there have been only two measurements of the ISRF in the LUV range.  These were done with the ultraviolet spectrographs (UVS) on the Voyager 1 and 2 missions and the Far Ultraviolet Spectrographic Explorer (FUSE).  Murthy and Sahnow compiled the results between FUSE and UVS and found a strong correlation of LUV flux to the \SI{100}{\micro\meter} thermal flux of the interstellar dust grains \cite{Murthy04}.  Interestingly, the variation was stronger than had been measured by previous studies in FUV wavelengths due to the higher optical depth of dust grains to LUV light. Their review of the diffuse UV was limited in solid angle, so it is less reliable to determine an intensity range using only the FUSE  data (1600 -- $\sim$ 3 $\times$ $10^5$ continuum units\footnote{Continuum unit = photons cm$^{-2}$ s$^{-1}$ \AA$^{-1}$ sr$^{-1}$.}).  A subsequent study by Murthy (2014)\cite{Murthy14} of GALEX data provides a more useful estimate of the intensity range of dust that contributes to the ISRF which is $\sim$~400 -- 2000 continuum units from all-sky maps binned to 6$'$. The interaction of the ISRF with dust grains is the most direct method we have to study interstellar dust and can give valuable information on the interstellar dust grain parameters such as albedo and the forward scattering parameter.  

\paragraph{Molecular Hydrogen}
Fluorescent emission by molecular hydrogen (\ce{H2}) contributes to about 10\% of the ISRF.  \ce{H2} comprises about 20\% of all of the hydrogen in the Galaxy and plays an important role in star formation and the cooling of interstellar clouds.\cite{Jo17, Shull82}  It is a homonuclear molecule lacking a dipole moment, which makes it difficult to observe because transitions between different vibrational and rotational states within the ground state are forbidden \cite{Shull82}.  When \ce{H2} absorbs photons in the Lyman ({\bsig}) and Werner ({\cpi})bands shortward of \SI{1100}{\angstrom}, 10\% will dissociate while the other 90\% will fluorescently de-excite ({\bsig}$\rightarrow${\xsig}, {\cpi}$\rightarrow${\xsig}) through various UV transitions to ro-vibrational excited states of the ground state, which further fluoresce into the infrared.  It is through these measurements of \ce{H2} fluorescence that it can be observed directly if there is an excitation source nearby.  Jo et al. (2017)\cite{Jo17} sought to map \ce{H2} fluorescence across the entire Milky Way using data from the FIMS/SPEAR imaging spectrograph (with an imaging resolution of ~5' and a spectral resolution of ~550), which covered roughly 76\% of the sky.  While they were able to produce maps from the long wavelength channel (1350-\SI{1700}{\angstrom}), strong contamination from geocoronal Ly$\alpha$ emission lines made their short wavelength data (900-\SI{1150}{\angstrom}) unusable. 

\paragraph{Extragalactic Background Light}
The extragalactic background light (EBL) is the radiation across the entire electromagnetic spectrum coming from all sources from outside of our home Galaxy.\cite{Hill18, Murthy19}  In the ultraviolet the EBL mainly comes from star forming galaxies and high energy sources such as active galactic nuclei.  Using data from GALEX, Chiang et al. (2019)\cite{Chiang19} created a redshift dependent method to map the broadband intensity of the EBL in the FUV and found after foreground removal that the peak intensity of the EBL was ~89 continuum units.  However even after removing all known foregrounds they determined there was a 200 -- 500 continuum unit foreground that was from an unknown galactic source.

So far, the measurements of the cosmic UV background are so few there has been no attempt to calculate its power spectrum.  Figure~\ref{f2}a, compiled by Hill et al., demonstrates the accumulation of all measurements to date of the cosmic background in all bands with Figure~\ref{f2}b showing a close-up of the ultraviolet EBL.\cite{Hill18}  The purple block indicates the LUV region of the spectrum, which has no measurements at all.  This lack of measurements is due to extragalactic signals being so faint that they are easily overshadowed by instrument noise, absorbed by the interstellar medium completely, or overwhelmed by geocoronal light.  While there have been a few attempts to measure the EBL, there is a lot of controversy over whether or not these measurements are in fact the EBL or just unaccounted for noise.

Successful detection of the EBL will require a precise accounting of the intensity of dust scattered starlight and \ce{H2} fluorescence for foreground removal. A telescope with the ability to eliminate the geocorona and take photometric measurements as small as 300 continuum units would be required for a successful observation of the EBL in the LUV range.

\section{LUV Imager}

A means of filtering out Ly$\alpha$ in combination with a telescope outfitted with LiF/Al mirrors, and a solar-blind detector would enable the imaging of astrophysical sources in the LUV bandpass shortward of 1216 \AA\ and extending down to the LiF window transmission edge near 1040 Å \cite{Redwine13}.  Such a filter would open a new photometric survey capability for this wavelength regime, which because of the geocoronal intensity has primarily been explored spectroscopically and only rarely with imaging. Imaging and spectroscopy are complementary.  Imaging uses a bandpass filter to provide low spectral resolution observations over a wide angular field, while spectroscopy uses a dispersive optic to provide high spectral resolution observations over a narrow angular field.  Higher spectral resolution offers more chemical and kinematic information and a relative immunity to high monochromatic backgrounds, but with reduced sensitivity to faint objects and a requirement for longer integration times. 

By devising a means to filter out geocoronal Ly$\alpha$ we will enable LUV imaging surveys capable of efficiently discovering the faintest and rarest hot-stars, galaxies, and quasars whose LUV emissions provide a host of important astronomical diagnostics \cite{Sonneborn:2006}.  

\begin{figure}
    \centering
    \subfloat[\centering Efficiency]{{\includegraphics[height=6cm]{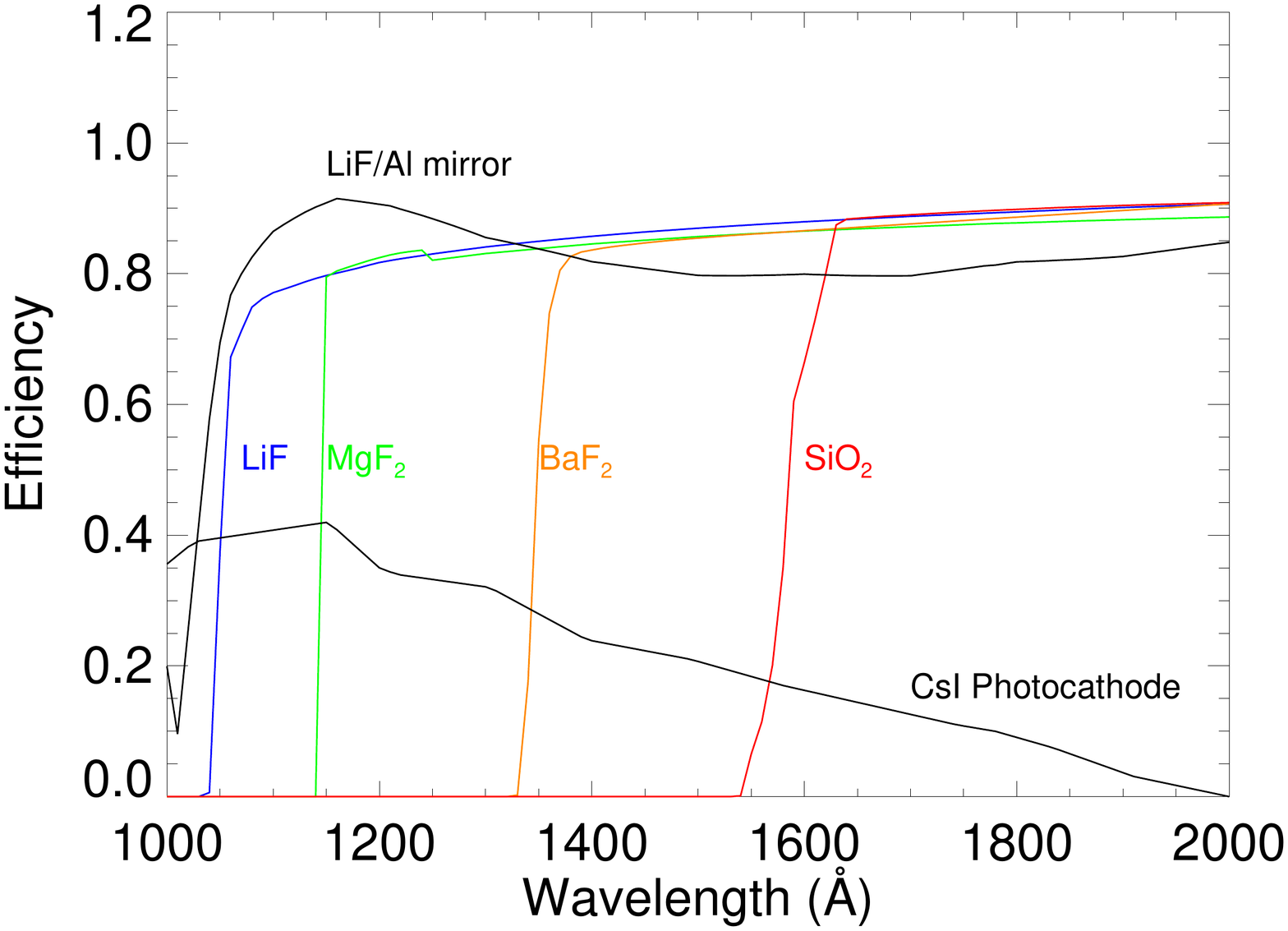} }}%
    \qquad
    \subfloat[\centering Effective Area]{{\includegraphics[height=6cm]{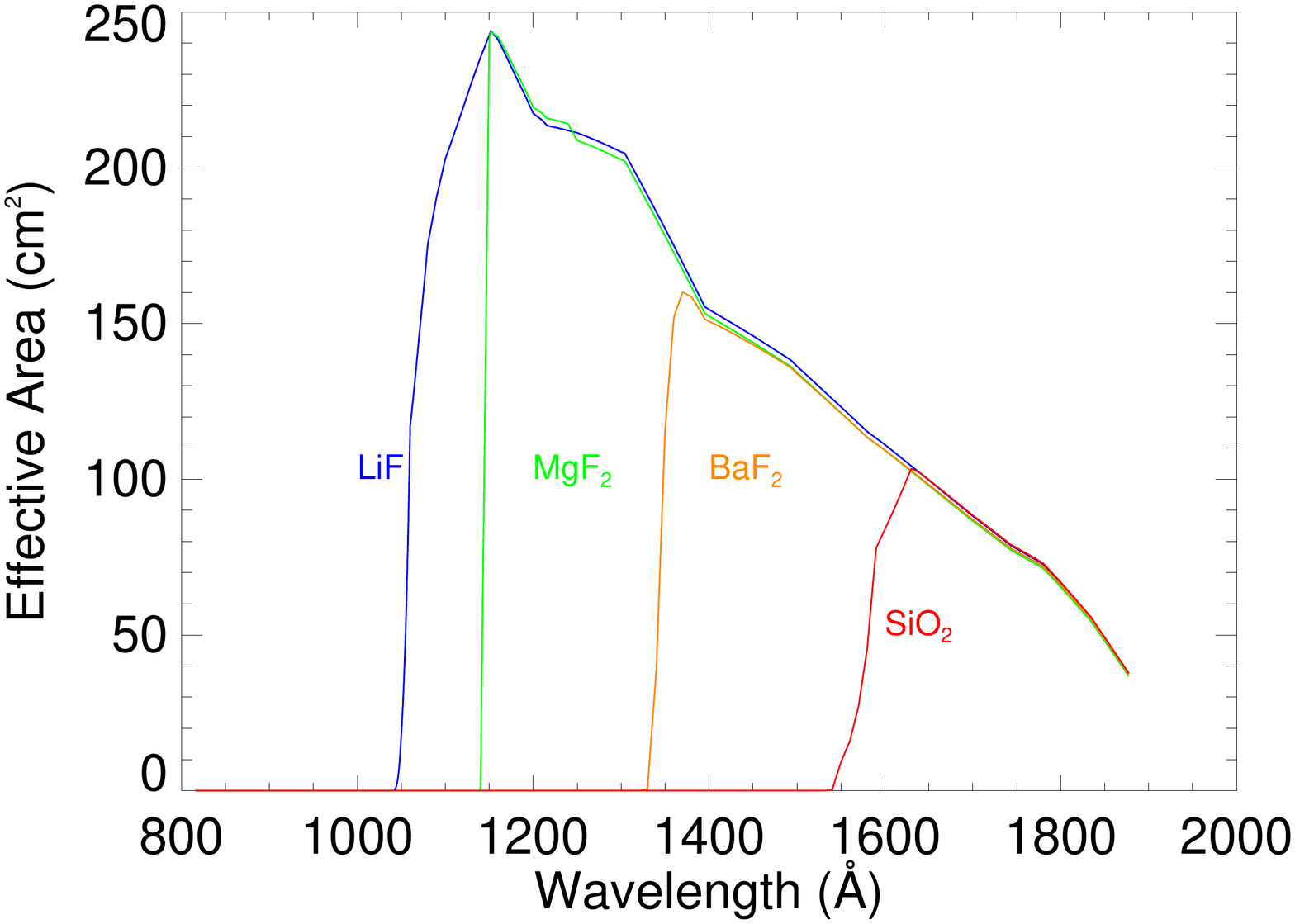} }}%
    \caption{\textbf{(a)} Transmission efficiencies for 1 mm thick filters. \textbf{(b)} Effective areas of long pass filters for a Schmidt style telescope with a 50 cm diameter LiF/Al mirror and a 30 cm diameter central obscuration.}%
    \label{f3}%
\end{figure}

As of now UV astronomers generally rely on algorithms, parameterized fits, or other noise subtraction techniques to get rid of the overwhelming signal of geocoronal Ly$\alpha$.  These methods are not proven to be completely effective because of the changing nature and extreme intensity of the geocorona.  Fortunately the means of selectively filtering monochromatic radiation exists in the form of a hydrogen absorption cell (H-Cell), wherein H$_2$ gas, sealed in a tube with transparent windows on each end, is dissociated into its neutral atomic form by exposure to a hot tungsten filament. The next generation of UV satellite missions would thus benefit from the use of an H-Cell as an atomically narrow band Ly$\alpha$ rejection filter, in combination with a variety of absorption-edge cutoff windows.  

\subsection{Photometry in the LUV Enabled by a Ly$\alpha$ Filter  }
\label{PhLUV}


Ultimately, we envision the development of a ``lensless'' Schmidt \cite{Fawdon:1989} style telescope (a single bounce telescope as suited to the photon starved farUV)  feeding a photon-counting multi-element 2-dimensional microchannel plate detector \cite{Siegmund:1993, Siegmund:2013} with an absorption cell to completely absorb the geocoronal Ly$\alpha$.  With Ly$\alpha$ eliminated we would use a set of nested longpass filters (windows) in between the absorption cell and the detector to define a photometry system similar to that used on the Solar-Blind Channel of the Advanced Camera for Surveys \cite{Ford:1998} on the {\it Hubble Space Telescope} ({\it HST}).

We can define a LUV photometric system using a process we refer to as differential long-pass filter imaging, which employs a set of windows with progressively shorter wavelength absorption edges (for example SiO$_2$; 160 nm, BaF$_2$; 135 nm,  MgF$_2$ 115 nm,  LiF; 103 nm) in front of a microchannel plate detector with a CsI photocathode.   By subtracting images acquired through adjacent long pass filters a bandpass can be synthesized.\footnote{A narrower set of bandpasses could be synthesized by inserting SrF$_2$; 125nm.}  For example, a system using LiF windows on the H-Cell with the addition of a SiO$_{2}$ edge cut-off filter would synthesize a bandwidth of 1050-\SI{1550}{\angstrom}. Figure~\ref{f3}a shows  the transmission efficiency for the edge filters along with the detector quantum efficiency (QE).  Figure ~\ref{f3}b shows the effective areas calculated using,

\begin{equation}
A_{eff} = A_{tel}R_{LiF/Al}(T_{LiF})^{2}T_{x}Q_{CsI} 
\end{equation}

\noindent Here, $A_{tel}$ is the area of the telescope, $R_{LiF/Al}$ is the reflectivity of the LiF/Al mirror, $T_{LiF}$ is the transmission efficiency of the LiF windows on the H-Cell, $T_{x}$ is the transmission efficiency of the edge filter, and $Q_{CsI}$ is the quantum efficiency of the detector. 
In Figure~\ref{f3}b we show the theoretical effective area for each channel, assuming a 50 cm diameter single bounce telescope with 30 cm central obscuration that has a LiF/Al mirror and a cell with two LiF windows.  The transmissions for each window (1mm thick) were calculated using the optical constants found in Palik (1985)\cite{Palik:1985} and subsequent editions. The CsI detective quantum efficiencies are typical of those described in McCandliss et al. (2010)\cite{McCandliss:2010} as provided by the Cosmic Origins Spectrograph (COS) detector group, and the LiF/Al mirror efficiencies were kindly provided by the GSFC mirror coating group \cite{Quijada:2017}.

\begin{figure}
\centering
\includegraphics[width=\textwidth,viewport= .5in .5in 9.5in 6.5in]{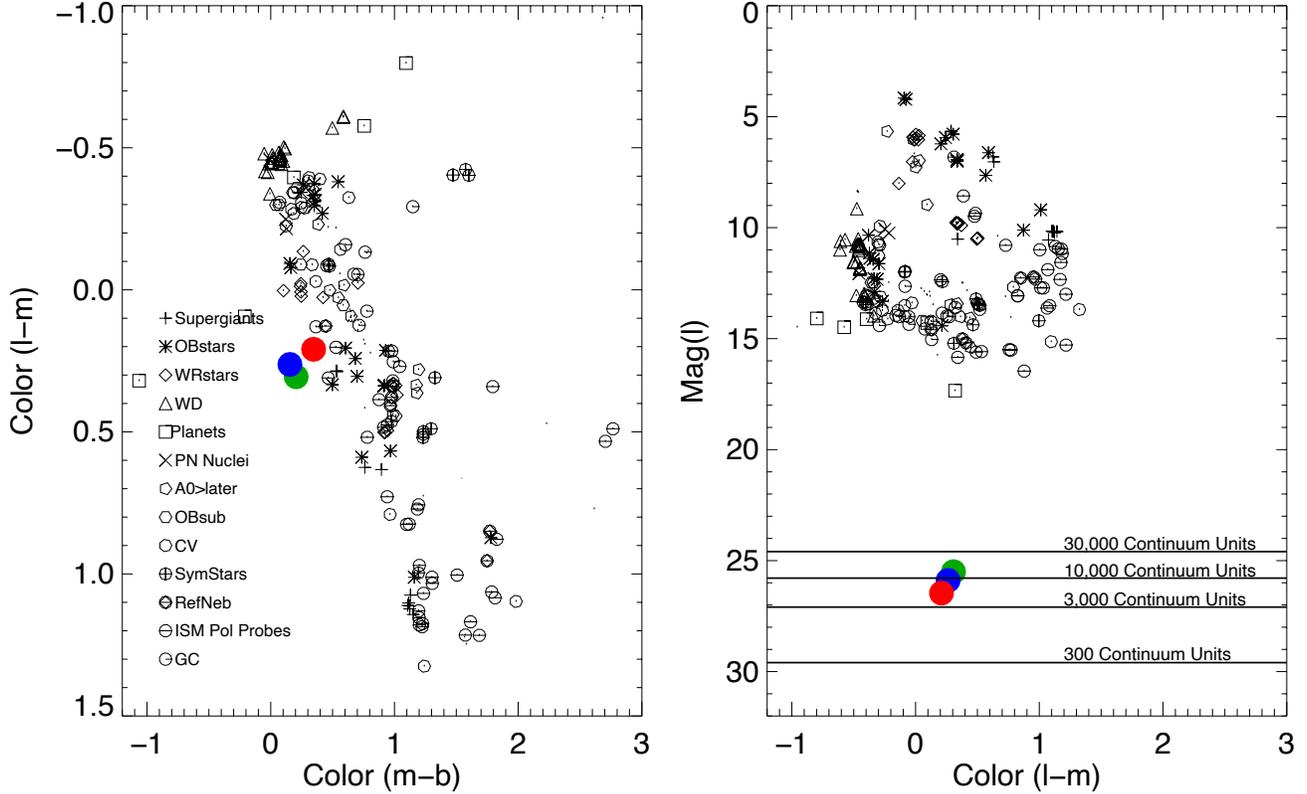}
\caption{Left - Color-color diagram  ($m-b$, $l-m$).  Right- Color-Magnitude (l) diagram for ($m-b$, $Mag(l)$). The green, blue and red dots denote where the fiducial Draine, Mathis and Habing models of the ISRF would fall within this context respectively.}
\label{f4}
\end{figure}

Differentially synthesized bands were defined by subtracting the adjacent long-pass effective areas shown in Figure~\ref{f3}b such that,
\begin{eqnarray}
A_l&=&A^{eff}_{LiF}-A^{eff}_{MgF_2} \\
A_m&=&A^{eff}_{MgF_2}-A^{eff}_{BaF_2} \\
A_b&=&A^{eff}_{BaF_2}-A^{eff}_{SiO_2} \\
A_s&=&A^{eff}_{SiO_2}. 
\end{eqnarray}
The mean flux within each synthesized narrow-band filter was then determined by integration over the bandpass.  The fluxes were converted to ab-magnitudes forming the colors $l-m$, $m-b$, and $b-s$.  The resulting color-color  ($m-b$, $l-m$) and color-magnitude ($m-b$, $l$) diagrams as derived from the {\it Hopkins Ultraviolet Telescope} ({\it HUT}) spectral database are shown in Figure~\ref{f4}.  The objects are all located within the Milky-Way. They represent various planets, hot stellar types, reflection nebulae and globular clusters.  An identification key for the different object types is included.  Also added into Figure~\ref{f4} are the three ISRF models produced by  Habing (1968)\cite{Habing68}, Draine (1978)\cite{Draine78}, and Mathis (1983)\cite{Mathis83}.

\begin{wraptable}{r}{.5\textwidth}
\renewcommand{\arraystretch}{1.3} 
\caption{LUV Photometry Parameters}
\vspace{4mm}
\centering
\begin{tabular}{|l|l|c|c|c|l|}
  \hline      
  \textbf{Band} & \textbf{$\lambda_{mean}^a$} & \textbf{$\lambda_{pivot}^a$} & \textbf{$\lambda_{rms}^a$} & \textbf{Sensitivity$^b$} \\
  \hline      
  \hline
$l$	&1125&1121&105&1.60\\
$m$	&1240&1239&126&3.82\\
$b$	&1455&1454&337&3.27\\
$s$	&1707&1705&588&2.69\\
  \hline  
\end{tabular}
\centerline{$^a$ \AA\ ,\ $^b$ $\times$ 10$^{15}$ (cnt s$^{-1}$)/(ergs $cm^{-2} s^{-1} \AA^{-1}$)}
\end{wraptable}

We have assessed the astronomical utility of this system using the spectroscopic database acquired by {\it HUT} during the Astro-1 and 2 missions \cite{Davidsen:1992}. {\it HUT} acquired spectra covering the FUV bandpass from 830 - 1880 \AA\ at a resolution $R \approx 1000$.  Spectra of good quality, having global count-rates in excess of 40 Hz were downloaded from the Mikulski Archive for Space Telescopes (MAST); $\approx$ 220 in all.   We removed emissions from Ly$\alpha$, \ion{N}{1} \lam 1200, and \ion{O}{1} \lam 1303, and \ion{O}{1} \lam 1356 under the assumption that all the observations would be acquired during orbital-night and through the Ly$\alpha$ absorption cell.

In Table 1 we list the synthesized band mean wavelength, pivot wavelength, rms wavelength, and sensitivity.  By way of example; if a star had a mean continuum flux in the $l$ band of 10$^{-15}$ ergs (an ab-mag$_{1121}$ = 19.8), then it would produce a count rate of 1.6 counts s$^{-1}$, which would yield a signal-to-noise ratio (S/N) of 4 in a mere 10 seconds.  A sounding rocket imaging experiment operating in the LUV with a 50 cm aperture, with the aforementioned parameter, and a typical integration time of $\approx$ 300 s could reach a S/N =3 on an object with a mean $l$ band flux of $\approx$ 2 $\times$ 10$^{-17}$ ergs; an ab-mag$_{1121}$ = 24.1.  It would provide the deepest LUV image ever acquired.  Such an instrument would be enormously useful for surveys to identify the most LUV intense objects for spectroscopic followup and to explore the LUV background.  

\begin{wraptable}{r}{.3\textwidth}
\renewcommand{\arraystretch}{1.3} 
\caption{Counts and signal-to-noise determined for the detection of 300 photons cm$^{-2}$ \AA$^{-1}$ s$^{-1}$ sr$^{-1}$ for a 2' field-of-view Schmidt style telescope with an integration time of 300 seconds.}
\vspace{4mm}
\centering
\begin{tabular}{|l|l|c|c|l|}
  \hline      
  \textbf{Filter}  & \textbf{Counts} & \textbf{S/N} \\
  \hline      
  \hline
$l$&557&7\\
$m$&1315&19\\
$b$&1004&20\\
$s$&723&27\\
  \hline  
\end{tabular}
\end{wraptable}

\begin{figure}
\centering
\includegraphics[height=7cm]{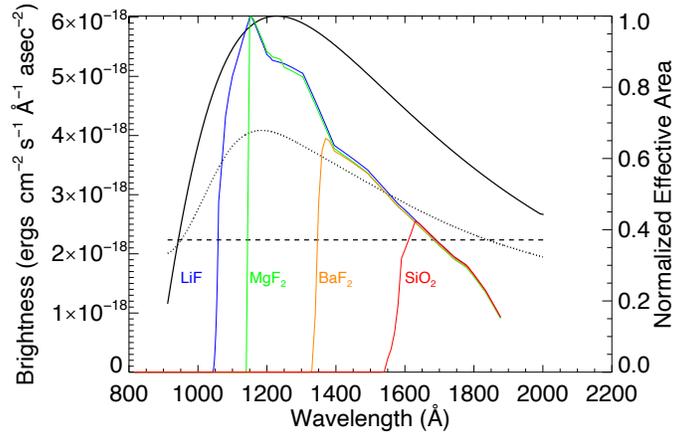}
\caption{Brightness of ISRF models overlaying the normalized effective areas shown in Figure~\ref{f3}b.  The filled line is Draine's (1978) model\cite{Draine78}, the dotted line is the model by Habing (1968)\cite{Habing68} and the dashed line is the model by Mathis (1983)\cite{Mathis83}.}
\label{f5}
\end{figure}

Table 2 lists the filtering scheme of the synthetic bandpasses for a wide-field (2 arcmin) detection of 300 photon cm$^{-2}$ \AA$^{-1}$ s$^{-1}$ arcsec$^{-1}$.  With these results we could perform the first ever detection of the ISRF in the LUV including but not limited to the very faint EBL.   Figure~\ref{f5} demonstrates where the three ISRF theoretical models by Draine, Habing, and Mathis would fall within the proposed filtering structure.

\section{Prototype Hydrogen Absorption  Cell}

\begin{figure} [ht]
    \centering
    \subfloat[\centering Hydrogen Cell]{{\includegraphics[height=6cm, width=5cm]{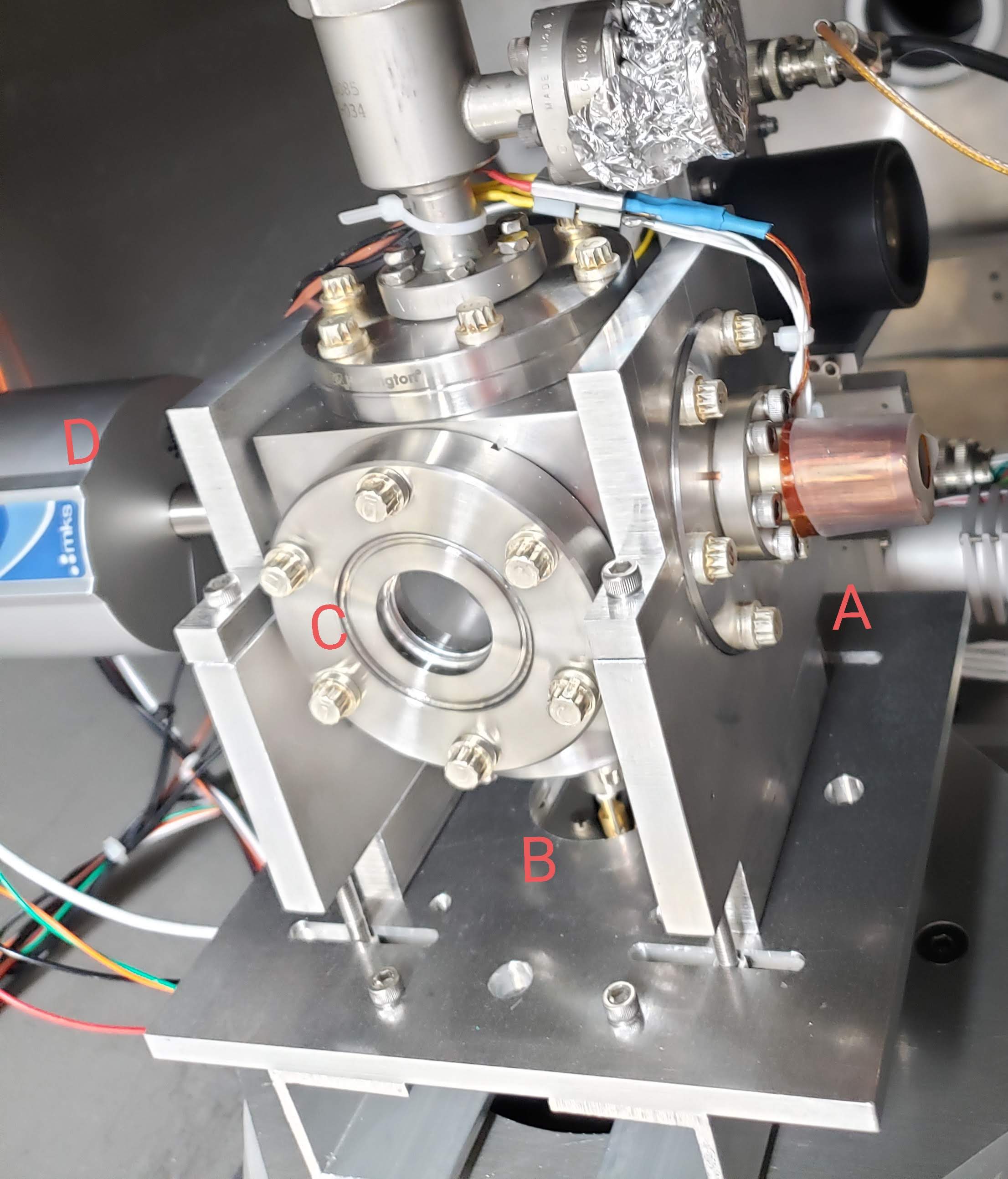} }}%
    \qquad
    \subfloat[\centering Transmission]{{\includegraphics[height=6cm, width=8cm]{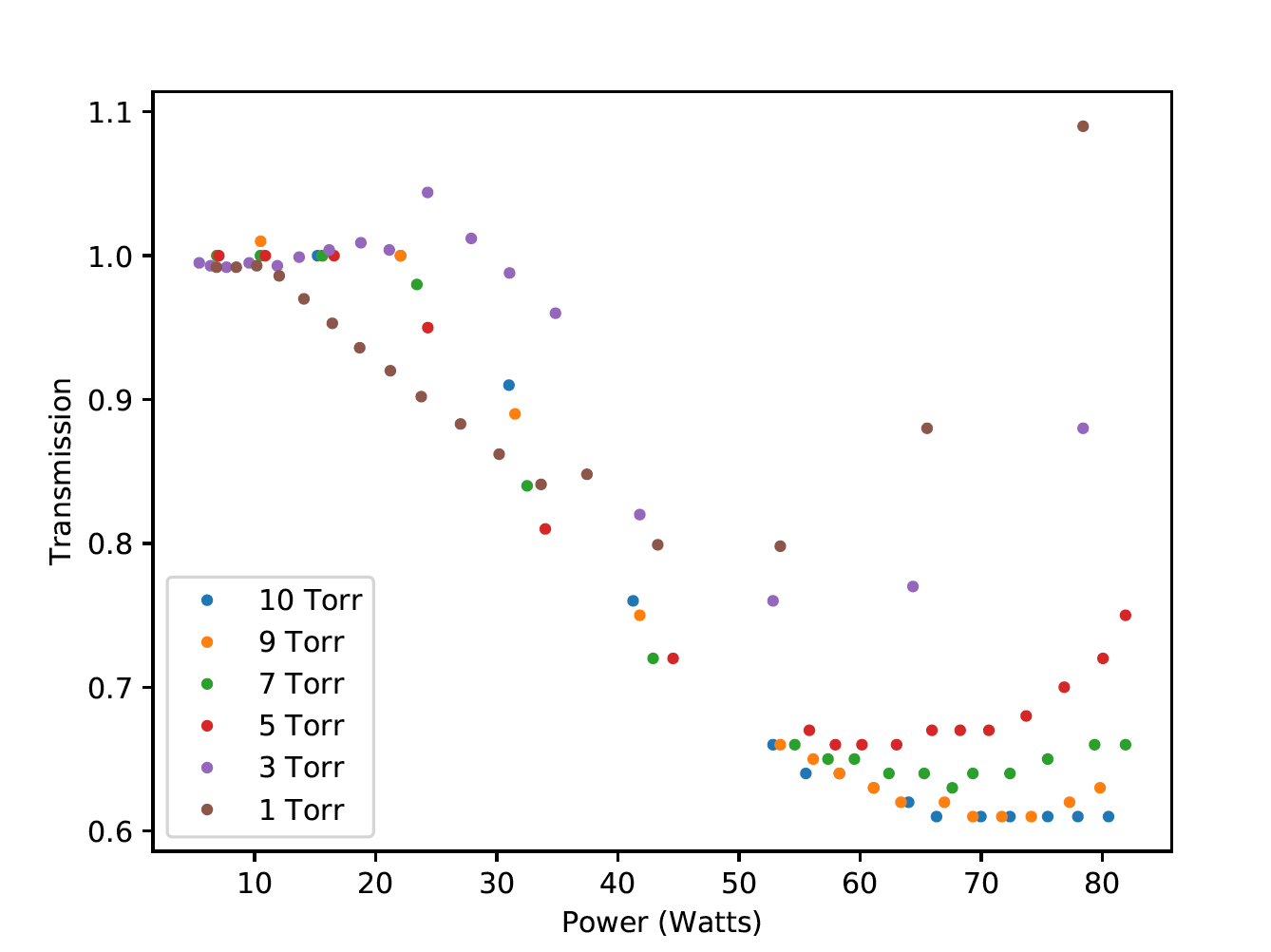} }}%
    \caption{\textbf{(a)} Hydrogen Absorption Cell with components labeled.  A) non-evaporable getter with heater attached B) wiring for tungsten filament C) lithium fluoride window D) pressure gauge. \textbf{(b)} Transmission of Ly$\alpha$ through the hydrogen cell at varying wattages and pressures.}%
    \label{f6}%
\end{figure}

The prototype H-Cell currently being tested in the lab is a small low-pressure chamber sealed between a pair of MgF$_{2}$ windows, allowing transmission down to \SI{1150}{\angstrom}.  A non-evaporable reversible getter is saturated with molecular hydrogen, which it is able to then outgas with the help of a small heater.  The getter heater pairing allows us to manipulate the pressure of the \ce{H2} in the cell so that we can perform transmission tests at various column densities.  Once the pressure of molecular hydrogen is at the desired value, it is converted to its neutral atomic form with the use of a hot tungsten filament.  Figure~\ref{f6}a shows the different components of the hydrogen absorption cell.

We mounted the H-Cell into a vacuum chamber where it was then illuminated with an UV monochromator dialed to emit at 1216~\AA.  The transmission data was taken by a G-tube photo-multiplier sensitive to Ly$\alpha$.  Initial tests were done to first determine the best wattage to apply to the tungsten filament in order to dissociate the most \ce{H2}.  With our current setup, it was found to be  $\sim$65 Watts.  We tested the filament in a pressure range of 1 to 10 torr to see where the transmission of Ly$\alpha$ would attenuate the most.  Figure~\ref{f5}b demonstrates these preliminary results.  We reduced the transmission by 40$\%$ at the highest pressure of 10 torr; short of our goal of 100\%.  We have identified two possible mechanisms that have prevented the H-Cell from reaching complete absorption, electron impact excitation of \ce{H2} and \ce{HI}, and the high resistivity of the tungsten filament.

A notable feature in Figure~\ref{f6}b is the reversal in transmission of the lower pressure experiments (1, 3 and 5 torr) once a threshold power was achieved.  For example, the data taken at 1 torr shows the transmission dropping by about 20\% at 40-50 watts but then rising around 80 watts until the transmission becomes higher than 1.  We suspect, based on the work of Liu et al. (1998)\cite{Liu98} and James et al. (1997)\cite{James97}  that this could be due to the electron impact excitation of \ce{H2} and \ce{HI}, which has a threshold on the order of $\sim$10 eV.

It is vital that the rate of \ce{H2} dissociation far exceeds the rate of recombination in order for full absorption of Ly$\alpha$ radiation.  This makes the tungsten filament the most important part of a successful H-Cell.  When the equation for a blackbody is applied to the tungsten filament it becomes apparent that a filament with a smaller area should produce a higher temperature.  This is illustrated in Equation 6, where P is Power, A is area of the filament, $\sigma$ is the Stefan-Boltzmann constant, and T is temperature.  Based on the work of Smith and Fite the dissociation probability of \ce{H2} is close to zero at temperatures below 1800 K\cite{Smith62}.  When examining our current setup with a filament of $~$4 cm long at 65 Watts, the temperature according to Equation 1 would have been 1636 K.  This could explain the less significant attenuation of Ly$\alpha$ compared to the experiment done by Redwine (2013)\cite{Redwine:2013} who obtained full attenuation of Ly$\alpha$ with a similar setup but a smaller filament.  
Kuwabara et al.\cite{Kuwabara18} ran several tests on various sizes of tungsten filaments within an H-Cell and verified that the tungsten filaments with the smallest areas produced the best dissociation rates of \ce{H2}. 
\begin{equation}
P/A = \sigma T^{4} 
\end{equation}

The next step for this project is to run the experiment using a smaller tungsten filament.  Our new setup will have a filament that is 1.9 cm long and .04 cm wide with an area of .076 cm$^2$ this should allow us to reach much higher temperatures with a smaller power output ($\sim$4 watts).

\section{Summary}

Herein we have discussed how a Ly$\alpha$ filter would enable wide-field photometric observations of the ISRF in bandpasses that overlap that geocoronal line.  We present the concept of an LUV imager based a photometric system of bandpass filters synthesized by subtracting adjacent edge filter images.  Effective areas and other parameters for this photometric system are estimated, assuming a ``lensless’’ Schmidt camera with a 50 cm diameter and 30 cm central obscuration.    Color-color and color magnitude diagrams showing the loci of various galactic objects are provided along with the locations for various models of the ISRF.  We estimate that the good signal-to-noise could be achieved in each of the synthesized bands on measurements of the diffuse ISRF over a (2’)$^2$ solid angle at the 300 continuum unit level — as observed by {\it GALEX} at the galactic poles — in the 300 seconds of exoatmospheric time typically available to a sounding rocket.   Such an instrument would allow the first wide-field, broadband search for signs of the EBL in the wholly unexplored LUV region  shortward of Lyman alpha.

We also present the results of our prototype absorption cell experiments, which to date has provided a modest 40\% attenuation of Ly$\alpha$.  We present two possible reasons for this less than optimal performance: a voltage drop across the tungsten filament that exceed the threshold for electron excitation of \ion{H}{1} and/or H$_2$; and filament whose area was too large to reach a temperature high enough to effectively dissociate H$_2$.  Future work will incorporate a smaller filament, which is expected to yield total attenuation. 

Hydrogen cell technology is not a new concept.  It has been used on the SOHO spacecraft to detect solar Ly$\alpha$ and track comets, and has also been used on a few planetary missions to measure the exosphere densities of hydrogen and deuterium\cite{Kawahara97, Maki:1996, Bertaux95}.  As of now, however, it has never been employed for photometric applications in the LUV where it would make a major impact.  Such an instrument in combination with a synthetic edge filtering system will improve the quality of future LUV satellite missions and would clear the way for the first ever measurement of the ISRF in the LUV bandpass.  \\

\begin{center} {\bf ACKNOWLEDGEMENTS} \end{center}

This work is supported by NASA APRA grant NNX17AC26G to JHU  entitled, Rocket and Laboratory Experiments in Astrophysics.

\bibliography{report} 

\begin{thebibliography}{10}

\bibitem{Brinkmann69}
{Brinkmann}, R.~T., ``{Dissociation of water vapor and evolution of oxygen in
  the terrestrial atmosphere},'' {\em \jgr}~{\bf 74},  5355 (Jan. 1969).

\bibitem{Carruthers76}
Carruthers, G., Page, T., and Meier, R., ``Apollo 16 lyman alpha imagery of hte
  hydrogen geocorona,'' {\em Journal Of Geophysical Research}~{\bf 18},  10
  (1976).

\bibitem{Baliukin18}
Baliukin, I., Bertaux, J., Quemerais, E., Izmodenov, V., and Schmidt, W.,
  ``Swan/soho lyman-alpha mapping: The hydrogen geocorona extends well beyond
  the moon,'' {\em Journal of Geophysical Research: Space Physics}~{\bf 124},
  861--885 (2018).

\bibitem{Kameda17}
Kameda, S., Ikezawa, S., Sato, M., Kuwabara, M., Osada, N., Murakami, G.,
  Yoshioka, K., Yoshikawa, I., Taguchi, M., Funase, R., Sugita, S., Miyoshi,
  Y., and Fujimoto, M., ``Ecliptic north-south symmetry of hydrogen
  geocorona,'' {\em Geophysical Research Letters}~{\bf 44},  706--11 (2017).

\bibitem{Heays17}
{Heays}, A.~N., {Bosman}, A.~D., and {van Dishoeck}, E.~F.,
  ``{Photodissociation and photoionisation of atoms and molecules of
  astrophysical interest},'' {\em \aap}~{\bf 602},  A105 (June 2017).

\bibitem{Bowyer91}
Bowyer, S., ``The cosmic far ultraviolet background,'' {\em Annu. Rev. Astron.
  Astrophys.}~{\bf 29},  59--88 (1991).

\bibitem{Henry91}
Henry, R.~C., ``Ultraviolet background radiation,'' {\em Annu. Rev. Astron.
  Astrophys.}~{\bf 29},  89--128 (1991).

\bibitem{Murthy10}
Murthy, J., Henry, R., and Sujatha, N., ``Mapping the diffuse ultraviolet sky
  with the galaxy evolution explorer,'' {\em ApJ}~{\bf 724},  1389--1395
  (2010).

\bibitem{Murthy04}
Murthy, J. and Sahnow, D., ``Observations of the diffuse far-ultraviolet
  background with the far ultraviolet spectroscopic explorer,'' {\em ApJ}~{\bf
  615},  315--322 (2004).

\bibitem{Murthy14}
{Murthy}, J., ``{GALEX Diffuse Observations of the Sky: The Data},'' {\em
  \apjs}~{\bf 213},  32 (Aug. 2014).

\bibitem{Jo17}
Jo, Y.-S., Seon, K.-I., Edelstein, J., and Han, W., ``Far-ultraviolet
  fluorescent molecular hydrogen emission map of hte milky way galaxy,'' {\em
  ApJS}~{\bf 231},  21 (2017).

\bibitem{Shull82}
Shull, J. and Beckwith, S., ``Interstellar molecular hydrogen,'' {\em Annu.
  Rev. Astron. Astrophys.}~{\bf 20},  163--190 (1982).

\bibitem{Hill18}
Hill, R., Masui, K., and Scott, D., ``The spectrum of the universe,'' {\em
  Applied Spectroscopy}~{\bf 72},  663--688 (2018).

\bibitem{Murthy19}
{Murthy}, J., {Akshaya}, M.~S., and {Ravichandran}, S., ``{The Diffuse
  Ultraviolet and Optical Background: Status and Future Prospects},'' {\em
  arXiv e-prints} ,  arXiv:1909.05325 (Sept. 2019).

\bibitem{Chiang19}
{Chiang}, Y.-K., {M{\'e}nard}, B., and {Schiminovich}, D., ``{Broadband
  Intensity Tomography: Spectral Tagging of the Cosmic UV Background},'' {\em
  \apj}~{\bf 877},  150 (June 2019).

\bibitem{Redwine13}
{Redwine}, K., {McCandliss}, S.~R., {Fleming}, B.~T., and {Pelton}, R.,
  ``{Hydrogen cells as narrowband geo-coronal Lyman-alpha rejection filters for
  astrophysical photometry},'' in [{\em UV, X-Ray, and Gamma-Ray Space
  Instrumentation for Astronomy XVIII}{\nolinebreak\hspace{0.1em}]},
  {Siegmund}, O.~H., ed., {\em Society of Photo-Optical Instrumentation
  Engineers (SPIE) Conference Series} {\bf 8859},  88590P (Sept. 2013).

\bibitem{Sonneborn:2006}
{Sonneborn}, G., {Moos}, H.~W., and {Andersson}, B.-G., eds.,  [{\em
  {Astrophysics in the Far Ultraviolet: Five Years of Discovery with FUSE
  }}{\nolinebreak\hspace{0.1em}]}, {\em Astronomical Society of the Pacific
  Conference Series} {\bf 348} (June 2006).

\bibitem{Fawdon:1989}
{Fawdon}, P. and {Gavin}, M.~V., ``{A Lensless Schmidt Camera},'' {\em Journal
  of the British Astronomical Association}~{\bf 99},  292--295 (Dec. 1989).

\bibitem{Siegmund:1993}
{Siegmund}, O.~H., {Gummin}, M.~A., {Stock}, J.~M., {Marsh}, D.~R., {Raffanti},
  R., and {Hull}, J., ``{High-resolution monolithic delay-line readout
  techniques for two-dimensional microchannel plate detectors},'' in [{\em
  Proc. SPIE Vol. 2006, p. 176-187, EUV, X-Ray, and Gamma-Ray Instrumentation
  for Astronomy IV, Oswald H. Siegmund; Ed.}{\nolinebreak\hspace{0.1em}]},
  {\bf 2006},  176--187 (Nov. 1993).

\bibitem{Siegmund:2013}
{Siegmund}, O.~H.~W., {Richner}, N., {Gunjala}, G., {McPhate}, J.~B.,
  {Tremsin}, A.~S., {Frisch}, H.~J., {Elam}, J., {Mane}, A., {Wagner}, R.,
  {Craven}, C.~A., and {Minot}, M.~J., ``{Performance characteristics of atomic
  layer functionalized microchannel plates},'' in [{\em Society of
  Photo-Optical Instrumentation Engineers (SPIE) Conference
  Series}{\nolinebreak\hspace{0.1em}]},  {\em Society of Photo-Optical
  Instrumentation Engineers (SPIE) Conference Series} {\bf 8859},  0 (Sept.
  2013).

\bibitem{Ford:1998}
{Ford}, H.~C., {Bartko}, F., {Bely}, P.~Y., {Broadhurst}, T., {Burrows}, C.~J.,
  {Cheng}, E.~S., {Clampin}, M., {Crocker}, J.~H., {Feldman}, P.~D.,
  {Golimowski}, D.~A., {Hartig}, G.~F., {Illingworth}, G., {Kimble}, R.~A.,
  {Lesser}, M.~P., {Miley}, G., {Neff}, S.~G., {Postman}, M., {Sparks}, W.~B.,
  {Tsvetanov}, Z., {White}, R.~L., {Sullivan}, P., {Krebs}, C.~A., {Leviton},
  D.~B., {La Jeunesse}, T., {Burmester}, W., {Fike}, S., {Johnson}, R.,
  {Slusher}, R.~B., {Volmer}, P., and {Woodruff}, R.~A., ``{Advanced camera for
  the Hubble Space Telescope},'' in [{\em Space Telescopes and Instruments
  V}{\nolinebreak\hspace{0.1em}]},  {Bely}, P.~Y. and {Breckinridge}, J.~B.,
  eds., {\em \procspie} {\bf 3356},  234--248 (Aug. 1998).

\bibitem{Palik:1985}
{Palik}, E.~D.,  [{\em {Handbook of optical constants of
  solids}}{\nolinebreak\hspace{0.1em}]} (1985).

\bibitem{McCandliss:2010}
{McCandliss}, S.~R., {France}, K., {Osterman}, S., {Green}, J.~C., {McPhate},
  J.~B., and {Wilkinson}, E., ``{Far-Ultraviolet Sensitivity of the Cosmic
  Origins Spectrograph},'' {\em \apjl}~{\bf 709},  L183--L187 (Feb. 2010).

\bibitem{Quijada:2017}
{Quijada}, M.~A., {del Hoyo}, J., {Boris}, D.~R., and {Walton}, S.~G.,
  ``{Improved mirror coatings for use in the Lyman Ultraviolet to enhance
  astronomical instrument capabilities},'' in [{\em Society of Photo-Optical
  Instrumentation Engineers (SPIE) Conference
  Series}{\nolinebreak\hspace{0.1em}]},  {\em Society of Photo-Optical
  Instrumentation Engineers (SPIE) Conference Series} {\bf 10398},  103980Z
  (Sept. 2017).

\bibitem{Habing68}
{Habing}, H.~J., ``{The interstellar radiation density between 912 A and 2400
  A},'' {\em \bain}~{\bf 19},  421 (Jan. 1968).

\bibitem{Draine78}
{Draine}, B.~T., ``{Photoelectric heating of interstellar gas.},'' {\em
  \apjs}~{\bf 36},  595--619 (Apr. 1978).

\bibitem{Mathis83}
{Mathis}, J.~S., {Mezger}, P.~G., and {Panagia}, N., ``{Interstellar radiation
  field and dust temperatures in the diffuse interstellar matter and in giant
  molecular clouds.},'' {\em \aap}~{\bf 500},  259--276 (Nov. 1983).

\bibitem{Davidsen:1992}
{Davidsen}, A.~F., {Long}, K.~S., {Durrance}, S.~T., {Blair}, W.~P., {Bowers},
  C.~W., {Conard}, S.~J., {Feldman}, P.~D., {Ferguson}, H.~C., {Fountain},
  G.~H., {Kimble}, R.~A., {Kriss}, G.~A., {Moos}, H.~W., and {Potocki}, K.~A.,
  ``{The Hopkins Ultraviolet Telescope - Performance and calibration during the
  Astro-1 mission},'' {\em \apj}~{\bf 392},  264--271 (June 1992).

\bibitem{Liu98}
{Liu}, X., {Shemansky}, D.~E., {Ahmed}, S.~M., {James}, G.~K., and {Ajello},
  J.~M., ``{Electron-impact excitation and emission cross sections of the
  H$_{2}$ Lyman and Werner systems},'' {\em \jgr}~{\bf 103},  26739--26758
  (Nov. 1998).

\bibitem{James97}
{James}, G.~K., {Slevin}, J.~A., {Shemansky}, D.~E., {McConkey}, J.~W., {Bray},
  I., {Dziczek}, D., {Kanik}, I., and {Ajello}, J.~M., ``{Optical excitation
  function of H(1s-2p) produced by electron impact from threshold to 1.8
  keV},'' {\em \pra}~{\bf 55},  1069--1087 (Feb. 1997).

\bibitem{Smith62}
{Smith}, Joe~N., J. and {Fite}, W.~L., ``{Reflection and Dissociation of
  H$_{2}$ on Tungsten},'' {\em \jcp}~{\bf 37},  898--904 (Aug. 1962).

\bibitem{Redwine:2013}
{Redwine}, K., {McCandliss}, S.~R., {Fleming}, B.~T., and {Pelton}, R.,
  ``{Hydrogen cells as narrowband geo-coronal Lyman-alpha rejection filters for
  astrophysical photometry},'' in [{\em UV, X-Ray, and Gamma-Ray Space
  Instrumentation for Astronomy XVIII}{\nolinebreak\hspace{0.1em}]},  {\em
  \procspie} {\bf 8859},  88590P (Sept. 2013).

\bibitem{Kuwabara18}
{Kuwabara}, M., {Taguchi}, M., {Yoshioka}, K., {Ishida}, T., {de Oliveira}, N.,
  {Ito}, K., {Kameda}, S., {Suzuki}, F., and {Yoshikawa}, I., ``{Evaluation of
  hydrogen absorption cells for observations of the planetary coronas},'' {\em
  Review of Scientific Instruments}~{\bf 89},  023111 (Feb. 2018).

\bibitem{Kawahara97}
Kawahara, T., Okano, S., Abe, T., Fukunishi, H., and Ito, K., ``Glass-type
  hydrogen and deuterium absorption cells developed for d/h ration measurements
  in the martian atmosphere,'' {\em Applied Optics}~{\bf 36},  10 (1997).

\bibitem{Maki:1996}
{Maki}, J., {Lawrence}, G., {Esposito}, L., {Lauche}, H., and {Ludwig}, M.,
  ``{The Cassini Hydrogen Deuterium Absorption Cell: A Remote Sensing
  Instrument for Atomic D/H Measurements at Titan},'' in [{\em AAS/Division for
  Planetary Sciences Meeting Abstracts \#28}{\nolinebreak\hspace{0.1em}]},
  {\em Bulletin of the American Astronomical Society} {\bf 28},  1132 (Sept.
  1996).

\bibitem{Bertaux95}
{Bertaux}, J.~L., {Kyr{\"o}l{\"a}}, E., {Qu{\'e}merais}, E., {Pellinen}, R.,
  {Lallement}, R., {Schmidt}, W., {Berth{\'e}}, M., {Dimarellis}, E.,
  {Goutail}, J.~P., {Taulemesse}, C., {Bernard}, C., {Leppelmeier}, G.,
  {Summanen}, T., {Hannula}, H., {Huomo}, H., {Kehl{\"a}}, V., {Korpela}, S.,
  {Lepp{\"a}l{\"a}}, K., {Str{\"o}mmer}, E., {Torsti}, J., {Viherkanto}, K.,
  {Hochedez}, J.~F., {Chretiennot}, G., {Peyroux}, R., and {Holzer}, T.,
  ``{SWAN: A Study of Solar Wind Anisotropies on SOHO with Lyman Alpha Sky
  Mapping},'' {\em \solphys}~{\bf 162},  403--439 (Dec. 1995).

\end{thebibliography}
\bibliographystyle{spiebib} 

\end{document}